\documentclass[showkeys,floatfix, noshowpacs, preprintnumbers, twocolumn, amsmath, amssymb, aps, superscriptaddress, prl, 10pt]{revtex4-1}

\usepackage{amsmath,amsthm,amssymb,amsfonts}

\usepackage{eqnarray}

\usepackage{graphicx}
\usepackage{multirow}
\usepackage[margin=1in]{geometry}
\usepackage{url}
\usepackage[usenames,dvipsnames]{xcolor}
\usepackage[breaklinks=true]{hyperref}
\usepackage{bbold}

\newtheorem{thm}{Theorem}

\newcommand{\ket}[1]{\left\vert{#1}\right\rangle}

\newcommand{\braket}[2]{\left\langle{#1}\vert{#2}\right\rangle}

\begin{document}


\title{Witnessing genuine multi-photon indistinguishability}

\author{Daniel J. Brod}
\affiliation{Perimeter Institute for Theoretical Physics, 31 Caroline Street North, Waterloo, ON N2L 2Y5, Canada}

\author{Ernesto F. Galv\~{a}o}
\affiliation{Instituto de F\'isica, Universidade Federal Fluminense, Av. Gal. Milton Tavares de Souza s/n, Niter\'oi, RJ, 24210-340, Brazil}

\author{Niko Viggianiello}
\affiliation{Dipartimento di Fisica, Sapienza Universit\`{a} di Roma,
Piazzale Aldo Moro 5, I-00185 Roma, Italy}

\author{Fulvio Flamini}
\affiliation{Dipartimento di Fisica, Sapienza Universit\`{a} di Roma,
Piazzale Aldo Moro 5, I-00185 Roma, Italy}

\author{Nicol\`o Spagnolo}
\affiliation{Dipartimento di Fisica, Sapienza Universit\`{a} di Roma,
Piazzale Aldo Moro 5, I-00185 Roma, Italy}

\author{Fabio Sciarrino}
\email{fabio.sciarrino@uniroma1.it}
\affiliation{Dipartimento di Fisica, Sapienza Universit\`{a} di Roma,
Piazzale Aldo Moro 5, I-00185 Roma, Italy}

\begin{abstract}

Bosonic interference is a fundamental physical phenomenon, and it is believed to lie at the heart of quantum computational advantage. It is thus necessary to develop practical tools to witness its presence, both for a reliable assessment of a quantum source and for fundamental investigations. Here we describe how linear interferometers can be used to unambiguously witness genuine $n$-boson indistinguishability. The amount of violation of the proposed witnesses bounds the degree of multi-boson indistinguishability, for which we also provide a novel intuitive model using set theory. We experimentally implement this test to bound the degree of 3-photon indistinguishability in states we prepare using parametric down-conversion. Our approach results in a convenient tool for practical photonic applications, and may inspire further fundamental advances based on the operational framework we adopt.

\end{abstract}

\maketitle

\textit{Introduction ---} Various notions of non-classicality, such as steering and nonlocality, are believed to offer different effective resources for quantum information processing \cite{Adesso16}. Similarly, quantum interference promises to enable several applications in quantum information, ranging from quantum communication \cite{Krenn16} to quantum computation \cite{Harrow17} and simulation \cite{Aspuru-Guzik12}.  A paradigmatic example of two-particle interference is the Hong-Ou-Mandel (HOM) effect in a balanced beam splitter \cite{Hong87}, where output events with a single photon per mode are strictly suppressed. In the many-particle setting, the model known as Boson Sampling \cite{AA10} results in computational advantage fuelled by quantum interference between $n$ photons evolving in a multi-mode linear-optical network. This promising model has resulted in a number of experiments implementing these devices \cite{Broome13boson,Spring13boson,Tillmann13boson, Crespi13boson, Spagnolo13birthday, Bentivegna15, Carolan15, Tillmann15, Wang_Pan17, Loredo17, He_Pan17, Wang18, Wang18time}. Following the recognition of the important role that quantum correlations play in quantum information protocols \cite{Dakic12}, various functionals have been proposed to quantify correlations \cite{Vedral97,Adesso16} or to testify the non-separability of specific quantum states \cite{Horodecki09}. Along the same lines, it is essential to witness and quantify multi-particle quantum interference as a resource to achieve quantum advantage.

Different techniques to discriminate indistinguishable bosons from distinguishable ones have been proposed \cite{Aaronson14,Tichy14,Aolita15,Crespi15,Walschaers16,Bentivegna2016,Wang2016,Dittel2017} and experimentally verified \cite{Spagnolo14,Carolan14,Bentivegna14,Crespi16,Viggianiello17optimal,Agresti17,Giordani18,Viggianiello18}, allowing the observation of genuine multi-particle quantum interference in specific scenarios. Moreover, it was recently suggested \cite{Agne17,Menssen15} that investigating intrinsically multi-particle interference beyond pairwise correlations is essential to obtain more complete knowledge of the full interference landscape. However, up to now no experimental procedure has been proposed to directly witness and quantify genuine $n$-photon indistinguishability.

In this Letter we propose a novel interferometric scheme capable of witnessing and quantifying genuine $n$-photon indistinguishability. We test a concrete implementation of the scheme in the form of a linear-optical bulk interferometer, which we use to characterize various indistinguishability regimes of three-photon states, reached by means of spectral filtering and temporal delays. Furthermore, we propose an intuitive set-theoretic model that captures quantitative features of genuine multi-photon indistinguishability. Our results show that it is possible to quantify true $n$-photon interference with relatively little experimental effort, using an interferometric setup suitable to miniaturisation using integrated photonic circuits \cite{Osellame2003,obrien09,szamait15,Carolan15,Tanzilli}, and which can be scaled up efficiently using known schemes \cite{barak, Crespi16}.

\textit{Partial photon distinguishability ---} Let us first consider the case of $n$ photons having some degree of indistinguishability, whose state can be described by the formalism developed in \cite{Rohde,Tichy,Scheschnovich}. In these papers, the starting point is a set of spectral distributions for the photons, which we denote as $\ket{\phi_{i}}$, such that two photons $i$ and $j$ are perfectly identical if $\braket{\phi_{i}}{\phi_{j}} = 1$ and distinguishable if $\braket{\phi_{i}}{\phi_{j}} = 0$. These distributions can carry information not only about actual spectra (i.e.\ frequency distributions), but also about spatial modes, polarization, and in general any photonic degree of freedom that is inaccessible to the detectors. As shown in \cite{Rohde,Tichy,Scheschnovich,Tillmann15}, any state in a regime of partial distinguishability can be written as a convex combination of extremal distinguishability regimes, where arbitrary subsets of photons are perfectly identical to each other but distinguishable from the rest. We denote these extremal density matrices by $\rho_{\vec{S}}$, where $\vec{S}$ is some string of letters that label identical photons. For example, $\rho_{AAA}$ labels a state of three identical photons, $\rho_{ABA}$ labels a state where the first and third photons are identical but perfectly distinguishable from the second, and so on.
\begin{figure}[t!]
\centering
\includegraphics[trim={1.6cm 0 1.5cm 0},clip,width=\linewidth]{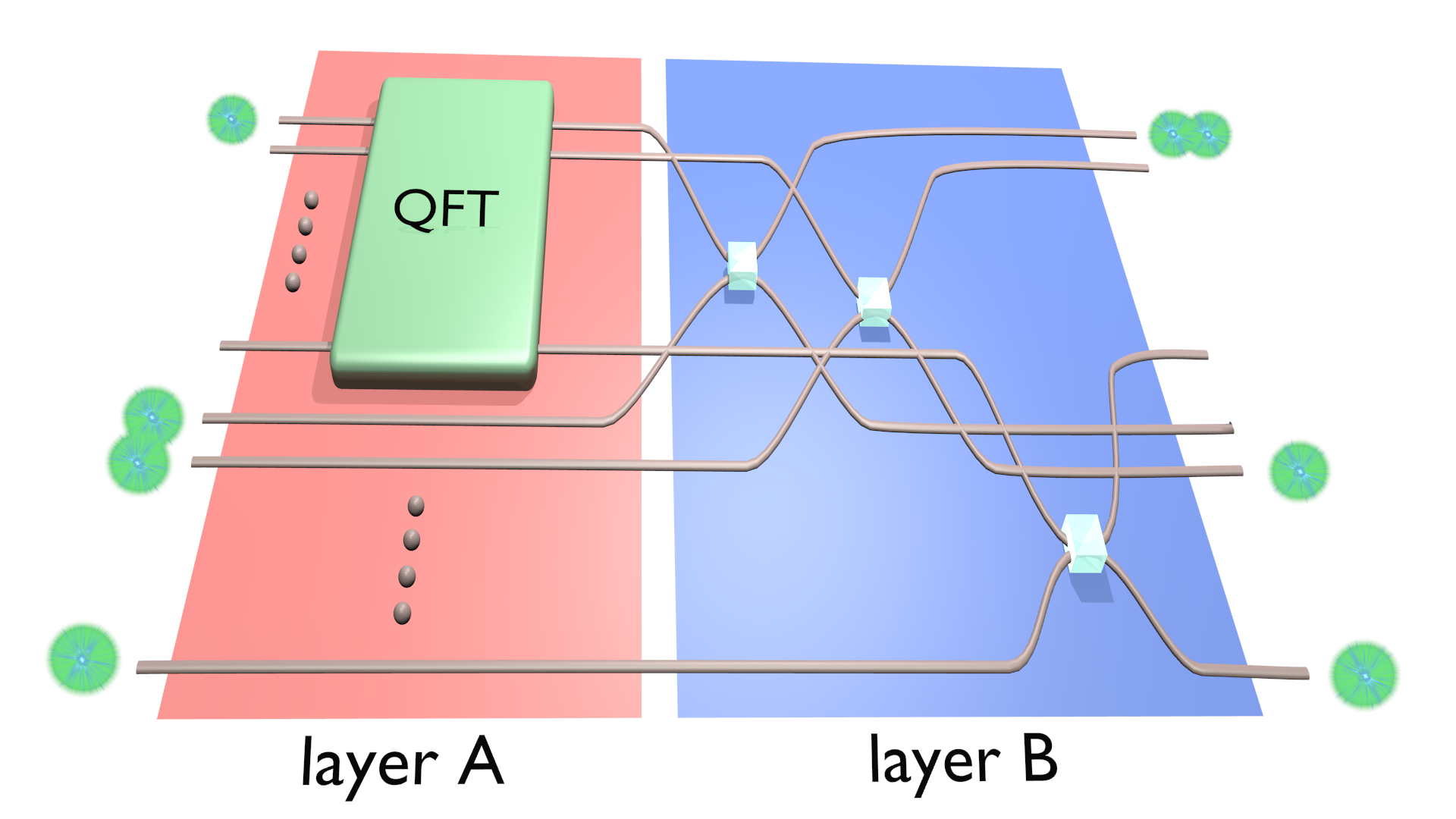}
\caption{Family of interferometers capable of witnessing genuine $n$-photon indistinguishability. QFT is the Quantum Fourier Transform acting on $(n-1)$ modes. Each output of the QFT is connected to a 50/50 beam splitter. As discussed in the main text, no mixture of extremal states with less than $n$ indistinguishable photons can result in a probability of bunching $p_b>\frac{2n-3}{2n-2}$. }
\label{fig:circuit}
\end{figure} 

Using this notation, we write the general $n$-photon state as
\begin{equation} \label{eq:convexn}
\rho = c_1 \; \rho_{AAA\ldots A} + \sum_i c_i \; \rho_{\vec{S_i}},
\end{equation}
where the sum runs over all possible decompositions of $n$ photons into subsets, with the exception of the state we denoted	 explicitly; naturally, $0 \leq c_i \leq 1$ and $\sum_i c_i =1$. In the formalism of \cite{Rohde,Tichy,Scheschnovich}, the $c_i$'s are subject to constraints coming from the assumption that each photon $i$ is in some pure state $\ket{\phi_{i}}$. Here we consider the more general scenario where the photons can be in an arbitrary convex combination (with non-unique decompositions, in general). We then propose the following criterion: a state $\rho$ (as in Eq. (\ref{eq:convexn})) displays genuine $n$-photon indistinguishability if $c_1>0$ in all of its convex decompositions.

Next we will describe a family of interferometers which can be used to experimentally demonstrate genuine $n$-photon indistinguishability. Our approach can be interpreted in an adversarial sense: if a source produces states with no support on the ideal $n$-photon state, i.e., if $c_1=0$ for some  decomposition of the state, it would fail the test no matter what mixtures it creates of the remaining extremal states.

\textit{Interferometers witnessing genuine $n$-photon indistinguishability ---} Consider the family of interferometers depicted in Fig.\ \ref{fig:circuit}, parametrized by the number of photons $n$.  These interferometers consist of a single $(n-1)$-mode discrete Quantum Fourier Transform (in layer $A$), followed by a sequence of  50/50 beam splitters connecting each QFT output mode with a different single mode from the bottom half of the interferometer (layer $B$). Note that all beam splitters in layer $B$ can be implemented in parallel. Intuitively, this interferometer is similar to a set of $n-1$ HOM tests in parallel, between the top input photon (which we call the reference photon) and every other.

\begin{figure}[b!]
\centering
\includegraphics[width=\linewidth]{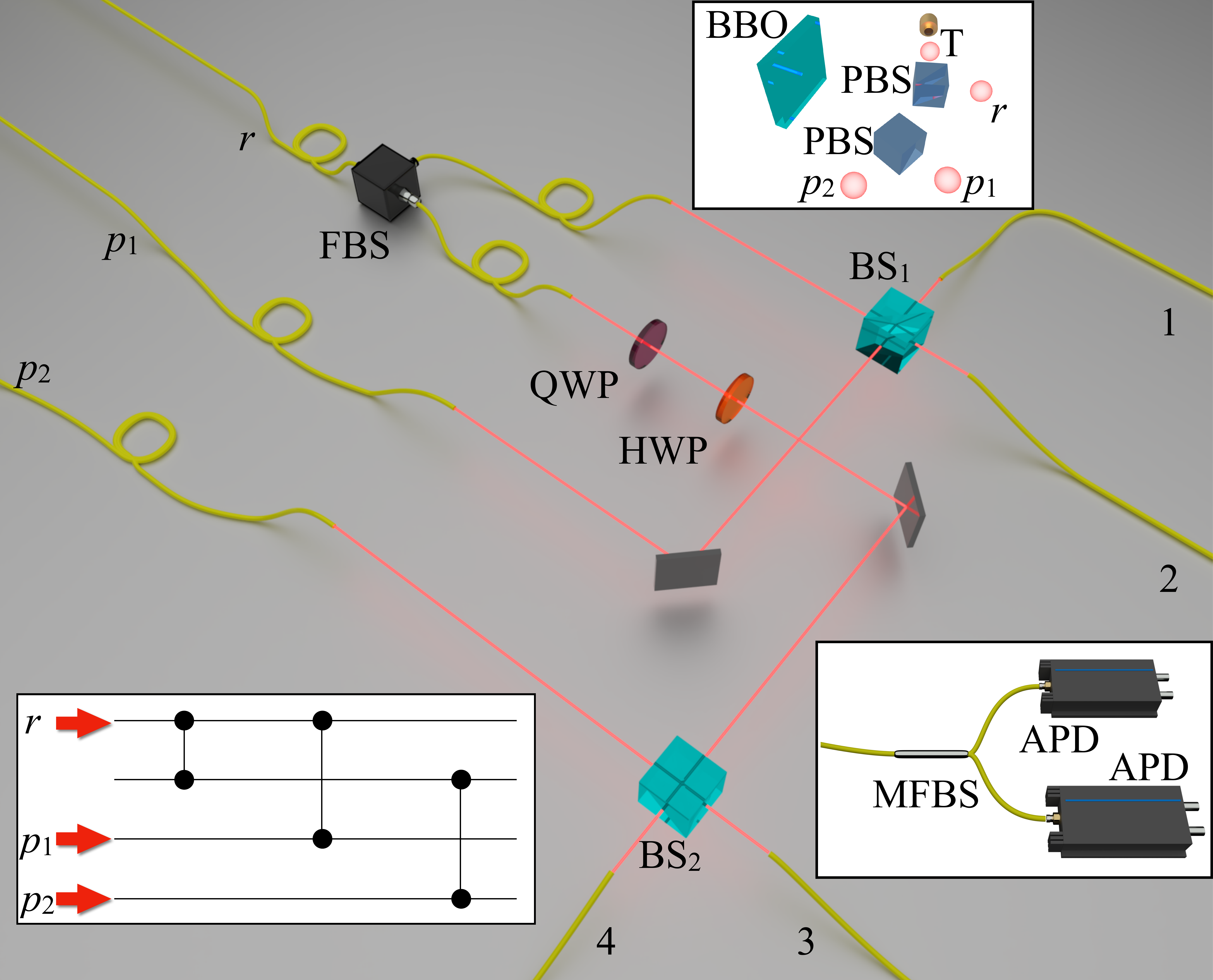}
\caption{Scheme of the experimental apparatus to test for genuine $3$-photon indistinguishability. Three-photon states are generated by a parametric down-conversion source operating in the double-pair emission regime (top right inset). Photons are made indistinguishable in polarization, spectrum and temporal delay, which was employed to control the degree of distinguishability. One photon ($r$) is split via a fiber beam splitter, and each output path interferes with one of the other photons in a second layer of beam splitters (see bottom left inset for a scheme of the circuit). Each output mode is measured by a detection block (bottom right inset) able to discriminate one- and two-photon contributions. Legend: BBO - Beta Barium Borate crystal, HWP - half-wave plate, PBS - polarizing beam splitter, FBS - fiber beam splitter, QWP - quarter-wave plate, BS - beam splitter, MFBS - multimode fiber beam splitter, APD - avalanche photodiode.}
\label{fig:expsetup}
\end{figure}

Acting on an input state of $n$ indistinguishable photons, our interferometer results in a probability of bunching $p_b=1$, due to the standard HOM interference of the reference photon in the top mode with one of the others. What is the probability of bunching for other extremal states? Note that the reference photon interferes with each of the bottom $(n-1)$ input photons with uniform probability $\frac{1}{(n-1)}$. Due to the HOM effect, if the two interfering photons are distinguishable (indistinguishable), the no-collision probability is $1/2$ (zero). Thus, each of the $(n-1)$ bottom photons that is distinguishable from the reference photon will result in a no-collision probability of $q=\frac{1}{2} \frac{1}{(n-1)}$. If $m$ of the bottom photons are distinguishable from the reference photon, the probability of bunching will be $p_b=1-mq=1-\frac{m}{2(n-1)}$. We see that the only extremal input state that achieves perfect bunching consists of exactly $n$ indistinguishable photons (i.e.\ $m=0$). Moreover, any convex combination of extremal states with fewer indistinguishable photons must have a bunching probability satisfying
\begin{equation}
p_b \le p_{*}\equiv\frac{2n-3}{2n-2}, \label{eq:witness}
\end{equation}
where we identify the threshold value $p_{*}$ for the bunching probability $p_b$. By picking different reference photons we actually obtain $n$ different inequalities of form (\ref{eq:witness}); a violation of any of them serves as a witness of genuine $n$-photon indistinguishability, that is, guarantees that $c_1>0$ in Eq.\ (\ref{eq:convexn}).

The amount of violation of any of our witnesses (\ref{eq:witness}) can be translated into bounds for the degree of genuine $n$-photon indistinguishability. To that end, assume first the worst-case scenario with a state of the type 
\begin{equation}
\rho = c_1 \; \rho_{AAA\ldots A} + (1-c_1) \; \rho_{ABA\ldots A},
\end{equation}
where $\rho_{ABA\ldots A}$ is some state that \emph{saturates} the bound of Eq.\ (\ref{eq:witness}). The probability of bunching $p_b$ is then: 
\begin{equation}
p_b= c_1 + (1-c_1)\; p_* \label{eq:worstcase}
\end{equation}
This gives a lower bound on the value of $c_1$. An analogous argument using the best-case scenario where photons are either all identical or all distinguishable leads to an upper bound. These together give the following range of values for $c_1$:
\begin{align}
\frac{p_b-p_*}{1-p_*} \leq c_1 \leq 2 p_b-1, \label{eq:c1fromp}
\end{align}
Thus, the amount of violation $p_b-p_*$ quantifies the amount of genuine $n$-photon indistinguishability. Note also that asymptotically (in $n$) we have $p_* = 1 - \textrm{O}(1/n)$. This means that, although our test becomes more stringent as $n$ increases, it is also efficient, in the sense that we expect to be able to resolve the gap between $p_b$ and $p_*$ with a number of events that scales like a polynomial in $n$.

\bigskip

\begin{figure*}[ht!]
\centering
\includegraphics[trim={3cm 0 3cm 2cm},clip,width=\textwidth]{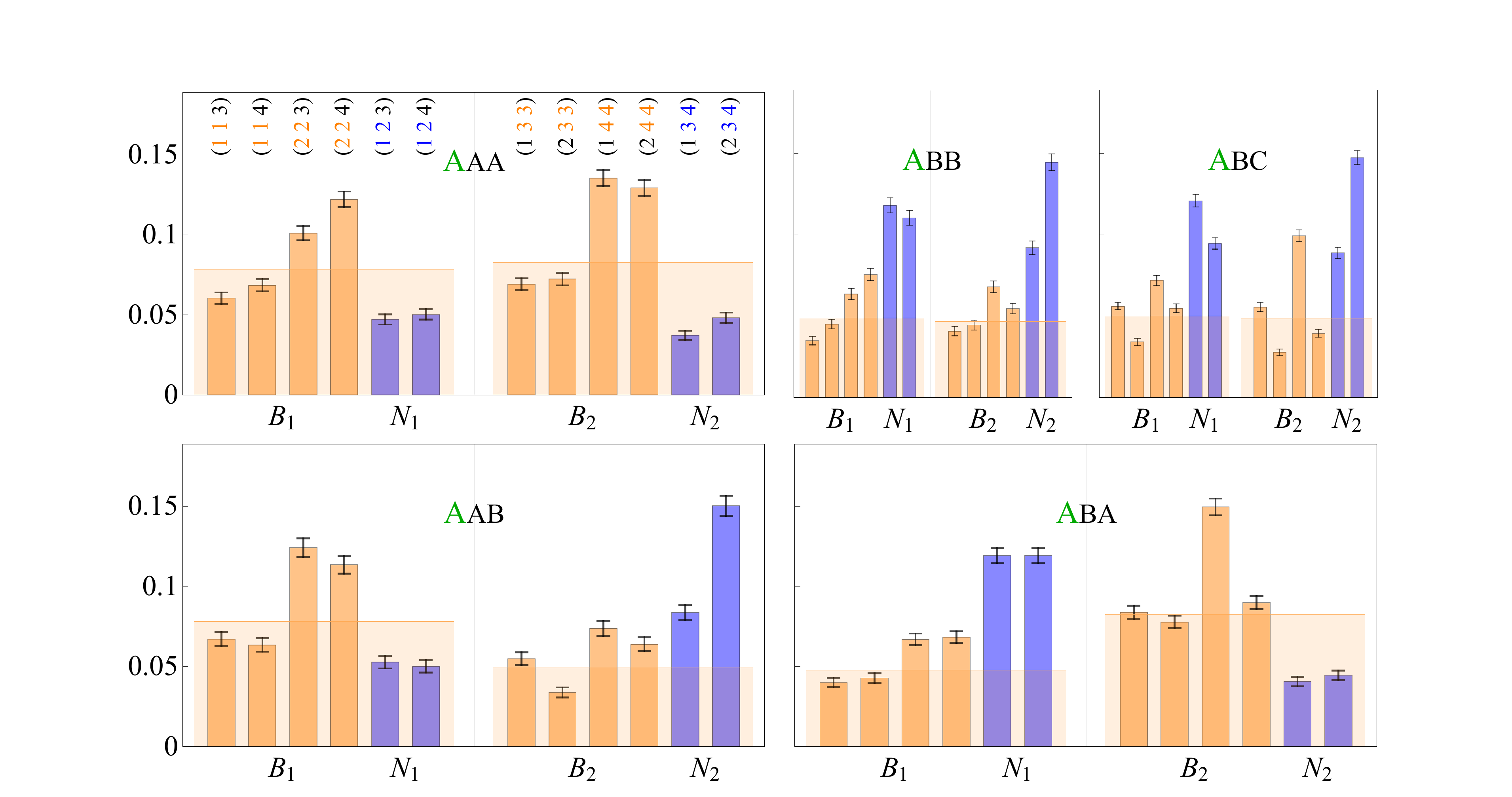}
\caption{Histograms of observed probability distributions over outputs for different distinguishability regimes (top of each panel), obtained via relative delays between input photons.  In all histograms events are ordered first by output beam splitter (BS$_{1}$, modes 1 and 2: left; BS$_{2}$, modes 3 and 4: right) and then by type of event (bunching: orange $B$; no-bunching: blue $N$). Bars are sorted following the mode assignment list shown in panel AAA, with mode numbers colored according to type of event. For each BS, orange shaded regions indicate the bunching probability $p_b^{(\mathrm{BS})}$ (rescaled by 0.1 to fit the plot). 
$p_b$ for BS$_{1}$ and BS$_{2}$ helps identify the distinguishability between their photons and the reference one, i.e. $p_b^{(\mathrm{BS}_1)} \sim p_b^{(\mathrm{BS}_2)}$ for AAA, ABB, ABC, $p_b^{(\mathrm{BS}_1)}>p_b^{(\mathrm{BS}_2)}$ for AAB, $p_b^{(\mathrm{BS}_1)}<p_b^{(\mathrm{BS}_2)}$ for ABA.
Data are corrected by excluding events with $n>3$ photons due to PDC multi-pair emission (see Supplemental Material \cite{supp}).}
\label{fig:overlaps}
\end{figure*}

\textit{Interpretation in terms of set theory ---} We now present a simple and intriguing interpretation of coefficient $c_1$ in Eq. \ref{eq:convexn} in terms of set theory. Let us associate a set with each single-photon state, and associate the size of the pairwise intersection of two sets $|A \cap B|$ with $\left| \braket{A}{B} \right| ^2$, i.e.\ the probability that a photon in state $\ket{A}$ passes for a photon in state $\ket{B}$ \cite{footnote1}.
The sets used to model single-photon states then have size one, as $|A|=|A \cap A|=\left| \braket{A}{A} \right| ^2=1$ for any $A$.
Now let us take $n$ sets $A_i$ ($i=1,2, ... , n)$ to represent the states of $n$ photons, and consider the $(n-1)$ pairwise intersections of a single reference set (call it $A_r$) with all others. In the Supplemental Material \cite{supp} we prove the following theorem:

\begin{thm}
Consider $n$ sets $A_i$ ($i=1,2, \ldots, n$), each of size 1. For any $r \in \{1,2, \ldots, n \}$, define
\begin{equation}
\mathcal{I}_r = \sum_{j|j \neq r} |A_r \cap A_j|.
\end{equation}
Then, for all $r \in \{1,2, \ldots, n \}$, the size of the common intersection of all sets satisfies
\begin{equation}
2-n+\mathcal{I}_r \leq \left | \bigcap_{j=1}^n A_j \right | \leq \frac{1}{n-1} \mathcal{I}_r \label{eq:lbintersec}
\end{equation}
\end{thm}

Let us now interpret this theorem in the light of our proposed association between pairwise set intersections and two-photon overlaps. First, recall that in the two-photon HOM effect, the probability of bunching can be given in terms of the two-photon overlap as $p_b^{\mathrm{HOM}}(r,j)=(1+\left|\braket{r}{j}\right| ^2)/2$ \cite{Garcia-Escartin, footnote1}. Now note that $p_b$ for our interferometer is the average probability of bunching of $(n-1)$ independent HOM tests between the reference photon (label it $r$) and all the others. If we use this to rewrite the bounds of Eq. (\ref{eq:lbintersec}) in terms pairwise photon overlaps, we find that $\left|\cap_{j=1}^n A_j \right|$ satisfies exactly the same bounds as $c_1$ in eq. (\ref{eq:c1fromp}). This suggests we identify these two quantities, i.e.\ the size of the intersection $\left|\cap_{j=1}^n A_j \right|$ on the set-theoretic side of the argument with the value of $c_1$, characterizing genuine $n$-photon indistinguishability. Our intuitive, operational interpretation for pairwise set intersections leads us to a nontrivial but consistent operational interpretation  of $\left|\cap_{j=1}^n A_j \right|$ as well.

\begin{figure}[ht!]
\centering
\includegraphics[trim={1.225cm 0cm 1.2cm 0cm},clip,width=\linewidth]{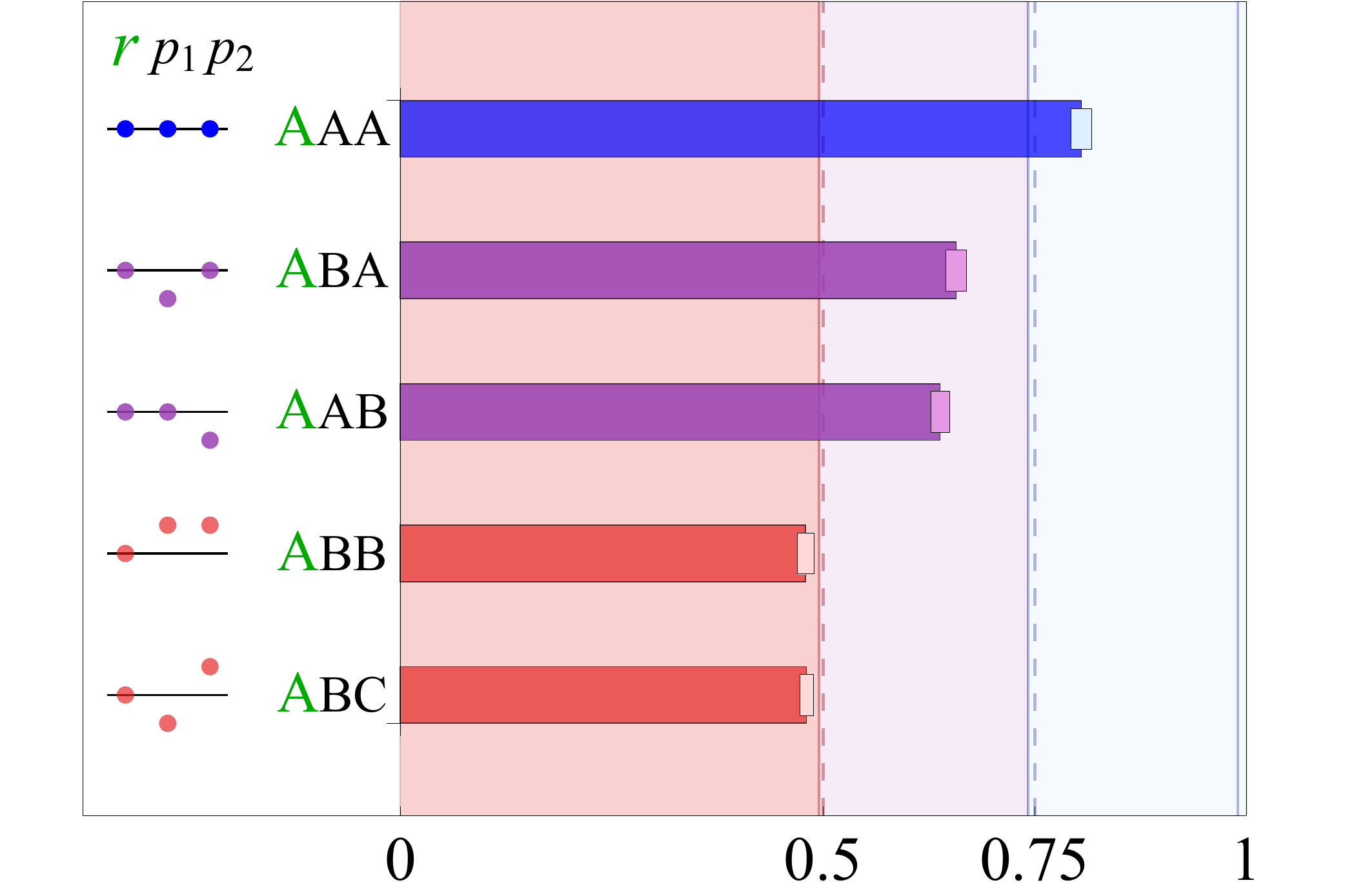}
\caption{Measured bunching probabilities $p_{b}$ for different regimes of distinguishability, i.e. when the reference photon $r$ is synchronized temporally with all the others ($p_1,p_2$) (blue), only one (purple) or none (red) (shown on the right). Red bars exhibit similar behaviours since the test does not discriminate the two cases. Purple bars are slightly different due to the overlap between photons belonging to the same or different pair. The dashed blue (red) vertical line is the threshold for indistinguishable (distinguishable) particles. Solid lines are the thresholds that account for the non-ideal unitary implementation (see Supplemental Material \cite{supp}).}
\label{fig:measures}
\end{figure}

\textit{Experimental implementation ---} We now describe the experimental scheme designed to witness and quantify genuine $3$-photon indistinguishability. To generate the 3-photon states we used 4-fold coincidence events from a parametric down-conversion (PDC) source. One of the four photons acts as a trigger, heralding the other three that are injected into the interferometer after proper synchronization. The interferometer is made up of two layers: layer $A$ consists of a single-mode in-fiber balanced beam splitter, while layer $B$ comprises two beam splitters, each connected to layer $A$ via one input port. A schematic representation of the setup is shown in Fig. \ref{fig:expsetup}, with further information in the Supplemental Material \cite{supp}. We measured the probability of bunching events $p_b$ for all five extremal states of 3 photons (Eq.\ \eqref{eq:convexn}), by introducing appropriate temporal delays and collecting up to $N \geq 2000$ events per state. In Fig.\ \ref{fig:overlaps} we report the full set of output measurements for all extremal states of distinguishability. In Fig.\ \ref{fig:measures} we report the measured values of $p_b$ for all extremal states. For the fully indistinguishable state, our measured witness was $p_b=0.805\pm 0.012$, which according to our bounds from Eq.\ \eqref{eq:witness} guarantees that $c_1$ must lie in the interval $0.22\pm 0.04 \leq c_1 \leq 0.61 \pm 0.02$, thus confirming genuine 3-photon indistinguishability. In fact, from the experimental data in Fig. \ref{fig:overlaps} we can obtain a tighter upper bound for $c_1$. It is easy to check that the probability of bunching of state (\ref{eq:convexn}) for beam-splitter BS$_{1}$ satisfies $2p_b^{\mathrm{HOM}}(\mathrm{BS}_1)-1=c_1+c_2$, where $c_2$ is associated with state $\rho_{AAB}$. Since all $c_j \ge 0$, $c_1 \le 2p_b^{\mathrm{HOM}}(\mathrm{BS}_1)-1$. An analogous reasoning for BS$_{2}$ gives $c_1 \le 2p_b^{\mathrm{HOM}}(\mathrm{BS}_2)-1$. We can read out $p_b^{\mathrm{HOM}}$ for BS$_{1}$ and BS$_{2}$ in Fig. \ref{fig:overlaps}a; the smaller of the two is $p_b^{\mathrm{HOM}}(\mathrm{BS}_1) = 0.78\pm 0.01$, which results in the tighter bound $c_1 \le 0.57\pm 0.02$. Incidentally, these tighter upper bounds for $c_1$ also have a precise counterpart in our proposed set-theoretical model (see Supplemental Material \cite{supp}). As expected, for all other scenarios the threshold $p_{*}$ for genuine $3$-photon indistinguishability is not violated (see Fig. \ref{fig:measures}).

In order to investigate in more detail how $p_b$ depends on degrees of freedom other than time-of-arrival, we reduced the spectral indistinguishability by removing the interferential filters at the source, while keeping the photons temporally synchronized. With this set-up, our measured probability of bunching was $p_b=0.66\pm 0.01$, well below the minimum threshold necessary to witness genuine multi-photon indistinguishability.

\textit{Discussion ---} Multi-particle interference is a key resource for metrology and optical quantum computation. In this Letter we tackle the problem of testing whether a photon source prepares states with genuine $n$-photon indistinguishability, as opposed to convex combinations of states in which effectively only a smaller number of photons interfere. Our scheme is simple, scalable and, when applied to single-photon states, can be implemented using only elementary linear-optical elements. We experimentally demonstrate the protocol by characterizing three-photon states in all extremal regimes of distinguishability. Furthermore, we propose a model for multi-boson indistinguishability in terms of set theory, a connection that we expect will foster further theoretical investigations. Ultimately, its relatively low technological requirements make the implementation of our test suitable for miniaturization using integrated photonic circuits, which encourages its adoption in the characterization of large-scale multi-photon applications.

\bigskip

\begin{acknowledgments}
\textbf{Acknowledgements.}

ERC-Starting Grant 3D-QUEST (3D-Quantum Integrated Optical Simulation; grant agreement no. 307783): \href{http://www.3dquest.eu}{3D-Quest}; H2020-FETPROACT-2014 Grant QUCHIP (Quantum Simulation on a Photonic Chip; grant agreement no. 641039): \href{http://www.quchip.eu}{QUCHIP}. Brazilian funding agency CNPq.

\end{acknowledgments}

\bigskip

\section{Supplemental material}

\textit{Theorem proof ---} In this supplemental section we prove the theorem about set intersections, referred to in the main text:

\textbf{Theorem:} Consider $n$ sets $A_i$ ($i=1,2, \ldots, n$), each of size 1. For any $r \in \{1,2, \ldots, n \}$, define
\begin{equation}
\mathcal{I}_r = \sum_{j|j \neq r} |A_r \cap A_j|.
\end{equation}
Then, for all $r \in \{1,2, \ldots, n \}$, the size of the common intersection of all sets satisfies
\begin{equation}
2-n+\mathcal{I}_r \leq \left | \bigcap_{j=1}^n A_j \right | \leq \frac{1}{n-1} \mathcal{I}_r \label{eq:lbintersecx}
\end{equation}

\begin{proof} Let us first prove the lower bound.  Label the elements of the chosen reference set $A_r$ (of size one) by a continuous index $\lambda \in (0,1)$. Define a set membership function $\mu_i(\lambda)$ associated with the set intersection $A_i \cap A_r$, indicating which elements $e_\lambda \in A_r$ are also in $A_i$:
	\begin{eqnarray}
	\mu_i(\lambda)=\left\{
	\begin{array}{ll}
	1 & \text{if } e_{\lambda} \in (A_i \cap A_r) \\
	0 & \text{if } e_{\lambda} \notin (A_i \cap A_r) \\
	\end{array} 
	\right.
	\end{eqnarray}
	The size of each intersection $A_i \cap A_r$ is determined by integration over $\lambda$ of the corresponding membership function:
	\begin{equation}
	|A_i \cap A_r|=\int_0^1 \mu_i(\lambda) d\lambda .
	\end{equation}
	
	The functions $\mu_{j}(\lambda)$ ($j\neq r$) can be used to obtain bounds on the size of the common intersection $\cap_{j=1}^n A_j$. An element $e_{\lambda}$ belongs to $\cap_{j=1}^n A_j$ iff it belongs to all $(n-1)$ pairwise intersections, i.e. iff $\sum_{j \neq r} \mu_{j}(\lambda)=n-1$. This means that the set membership function for $\cap_{j=1}^n A_j$ is $\max\left[0, \left(\sum_{j|j \neq r} \mu_{j}(\lambda)\right) -(n-2) \right]$. Integrating over $\lambda$, we obtain the size of the common intersection, and a simple lower bound for it:
	
	\begin{align}
	\left| \bigcap_{j=1}^n A_j \right |& =
	\int_0^1 \max \left[0, \left(\sum_{j|j \neq r} \mu_{j}(\lambda)\right) - (n-2) \right]d\lambda \notag\\
	& \ge \int_0^1 \left[ \sum_{j|j \neq r} \mu_{j}(\lambda)- (n-2) \right] d\lambda \notag\\
	& =\int_0^1 \left[\sum_{j|j \neq r} \mu_{j}(\lambda)\right] d\lambda -(n-2) \notag\\
	& =\sum_{j|j \neq r} \left[ \int_0^1 \mu_{j}(\lambda) d\lambda \right] -(n-2)
	\end{align}
	Recognizing the integrals in the last expression as the size of the pairwise intersections, we obtain
	\begin{equation}
	\left|\bigcap_{j=1}^n A_j \right |\ge 2-n+\sum_{j|j \neq r} |A_r \cap A_j|.
	\end{equation}
	To complete the proof, we need to obtain the upper bound in Eq.\ (\ref{eq:lbintersecx}). But that follows trivially from the fact that the $n$-fold intersection between the sets must be no larger than any pairwise intersection, and consequently must also be no larger than the average between all pairwise intersections:
	\begin{equation}
	\left|\bigcap_{j=1}^n A_j \right |\le \min_{j|j\neq r}\left( |A_r \cap A_j|\right)   \le \frac{1}{n-1} \sum_{j|j \neq r} |A_r \cap A_j|.
	\end{equation}
	This gives us the desired upper bound, and thus concludes the proof.
\end{proof}

\textit{Experimental details ---}
Single photons are generated at 785 nm with a type-II PDC process. A 2-mm long Beta Barium Borate (BBO) crystal is pumped with a 392.5 nm wavelength, 630 mW field, obtained by second harmonic generation from a 180 fs duration, 76 MHz repetition rate, Ti:Sa pulsed laser. Photons are spectrally filtered by means of 3 nm interferential filters and coupled into single-mode fibers. The indistinguishability regime can be obtained with a polarization compensation stage in-fiber and using delay lines to control relative delay between the paths. The delays are also employed to change the photon distinguishability throughout the experiment. Since the reference photon is distributed to all beam splitters in layer $B$ and each path must work independently, a further compensation stage (HWP, QWP and PBS) was used for one of the two arms. Collision events (i.e. outputs with more than one photon per mode) are measured by exploiting single-photon pseudo-number resolving (PNR) detectors in the collection stage. We employed low losses in-fiber multimode beam splitters (FBS) with splitting ratios distributed in the range (0.5-0.7) connected up to 8 avalanche photodiodes (APDs) with an auxiliary detector for the trigger photon. The measoured 4-fold events are then collected with two electronic devices (ID800 by IdQuantique) connected in parallel, driven by routines in LabView and C.

\textit{Partial distinguishability and analysis of multiphoton contribution of PDC sources on experimental data ---}
Differences between the extremal distinguishability states ABA and AAB depend on the partial distinguishability between photons belonging to different pairs.  More specifically, PDC sources based on broadband pulses show a reduced purity of the generated states due to the presence of spectral correlations between photon pairs. Such feature limits the quality of the interference between photons belonging to different pairs. To overcome this issue we employed interferential filters with $\Delta\lambda=3$ nm. 

Another relevant aspect to consider when using PDC sources is multi-photon emission, due to the probabilistic nature of the PDC process. Indeed, by increasing the pump power the probability of generating higher photon number terms cannot be neglected:
\begin{equation}
\ket{\Psi} \sim \ket{0,0}+g\, \ket{1,1}+ g^2 \, \ket{2,2}+ g^3 \, \ket{3,3},
\end{equation}
where $g$ is the nonlinear gain of the source.


To correct for multi-pair emission, we exclude the events forbidden by evolutions with only one photon per input mode of the interferometer, and which can be generated only by higher photon number terms. Let us call $N_b$ ($N_{nb}$) the sum of all bunching (non-bunching) events and $N_{mp}$ the sum of events due to higher-order emissions. Therefore, the total number of events is approximately $N \sim N_b+N_{nb}+N_{mp}$. By defining $P'_{nb}=N_{nb}/(N_b+N_{nb})$ as the number of total 3-photon non-bunching events that do not involve multi-pair emission, the correct values $P'_b$ reported in Fig. 3 of the main text are calculated as
\begin{equation}
P'_b=1-P'_{nb}=1-\frac{N_{nb}}{N_b+N_{nb}}.
\end{equation}
In our experiments, higher order terms comprise about $\sim 3\%$ of the collected events.

\textit{Witness bounds for implementations of interferometers with non-ideal reflectivities ---}
The bounds for the witness described in the main text were obtained by considering ideal beam splitters in the implementation for both layers A and B. Hence, both QFT and $B_j$ are built using cascades of ideal $50:50$ beam splitters. 

However, for non-ideal experimental implementations the actual values of the reflectivities ($R$) of the beam splitters must be taken into account. To include this aspect, new bounds are calculated by considering non-ideal beam splitters $BS(r)$ modeled as
\begin{equation}
BS(r)=\left(
\begin{array}{cc}
r & \sqrt{1-r^2}  \\
\sqrt{1-r^2}  & -r  \\
\end{array}
\right),
\label{BS_general}
\end{equation}
where $r=\sqrt{R}$. Let us now first discuss the case of the experiment reported in the main text. In this scenario, the $QFT_{2\times 2}$ of the first layer is a beam splitter with reflectivity $R_A=0.49$, while the beam splitters $B_j$ in layer B both have reflectivities $R_B=0.45$. We calculated the new bounds considering the new unitary transformation $U(R_i)$ that takes into account these parameters. For a general reflectivity $0\leq R\leq 1$, the bunching probability $P_b$ of having two photons injected in two different inputs of a $BS(r)$ is 
\begin{equation}
P_{b}^{\mathrm{ind}}(R)= 4\, (1-R) \, R \qquad P_{b}^{\mathrm{dis}}(R)= 2\, (1-R) \, R,
\label{nocoll_generalRef}
\end{equation}
where $P_{b}^{\mathrm{ind}}(R)$ and $P_{b}^{\mathrm{dis}}(R)$ are the case of indistinguishable and distinguishable particles respectively. It is worth noting that any $R\neq 0.5$  results in $P_{b}^{ind}<1$. By inserting the reflectivity values of our experimental setup, we obtain $P_{b}^{ind}(R_B)=0.99$. The new bounds are calculated by considering each beam splitter independently, and by weighting the second layer with the splitting ratios of layer A. As for the ideal case, the bound saturates when $n-1$ photons are indistinguishable and only one is distinguishable from the others. Therefore, the threshold considering the weights of layer A becomes 
\begin{equation}
\tilde{P_b}^*= R_A \, P_{b}^{ind}(R_{B_1}) + (1-R_A) \, P_{b}^{dis}(R_{B_2}).
\label{coll_generalRef}
\end{equation}
By averaging over all possible $\binom{n}{n-1}-1=2$ extremal distinguishability scenarios where just one photon is distinguishable from the reference photon, we obtain the value $\tilde{P_b}=0.742$. 

It is possible to generalize the witness bound also for interferometers corresponding to $n>3$. Let us call $\tilde{Q}_{p\times p}$, with $p=m/2$, the unitary matrix that approximates the QFT in layer A.  Let us define set $\sigma= \{AA\dots AB, AA\dots BA,\dots, AB\dots A\}$ as the set of all possible extremal states in which only a single photon is distinguishable from the reference photon. Finally, $(1,p+1,\dots,m)$ is the input state for the test and $\bar{R}_B=(R_1,\dots,R_{p})$ the set of non-ideal reflectivities for the beam splitters in layer B. Therefore, the threshold for higher dimensions can be calculated as a weighted sum of bunching probabilities for each $B(R_j)$, averaging over extremal states in set $\sigma$ is
\begin{equation}
\begin{aligned}
\tilde{P_b}^*(\bar{R}_B)&= \Big\langle \lvert \tilde{Q}_{j',1} \rvert^2 \; P_{b}^{\mathrm{dis}}(R_{j'}) +  \\
&+ \sum_{\substack{j=1\\ j\neq j'}}^{p} \, \lvert \tilde{Q}_{j,1} \rvert^2 \; P_{b}^{\mathrm{ind}}(R_j) \Big\rangle,
\end{aligned}
\label{general_bound}
\end{equation}
where $R_j$ is the $j^{\mathrm{th}}$ beam splitter in layer B, and $j'$ is the index of the beam splitter with the distinguishable photon. We now show that $P^*$ bound is maximum when the reflectivities correspond to the ideal case. Indeed, if we consider the terms in Eq. \eqref{nocoll_generalRef}, we observe that each beam splitter contribution is upper bounded by the ideal reflectivity. Thus, it follows that, given a positive sum of these contributions (where $\sum_{i=1}^{p}\, \lvert Q_{i,1} \rvert^2 =1$), $P^*(R_i)$ for each $R_i$ is maximized by ideal values: $R_i=0.5 \; \forall i$. Therefore, the value obtained with real beam splitters is upper-bounded by the threshold assuming perfect optical elements, so that a violation of the ideal threshold guarantees the presence of genuine $n$-photon indistinguishability.


\bigskip



\begin{thebibliography}{50}


\bibitem{Adesso16}
\bibinfo{author}{G. Adesso, T. R. Bromley, and M. Cianciaruso}, 
\newblock \bibinfo{title}{Measures and applications of quantum correlations}, 
\newblock \emph{\bibinfo{journal}{J. Phys. A: Math. Theor.}}  \textbf{\bibinfo{volume}{49}}, \bibinfo{pages}{473001}  (\bibinfo{year}{2016}).

\bibitem{Krenn16}
\bibinfo{author}{M. Krenn, M. Malik, T. Scheidl, R. Ursin, and A. Zeilinger}, 
\newblock \bibinfo{title}{Quantum communication with photons}, pp 455--482
(\bibinfo{year}{Springer, Cham, 2016}).


\bibitem{Harrow17}
\bibinfo{author}{A. W. Harrow and A. Montanaro}, 
\newblock \bibinfo{title}{Quantum computational supremacy},
\newblock \bibinfo{journal}{Nature}
\textbf{\bibinfo{volume}{549}}, 203–-209
(\bibinfo{year}{2017}).


\bibitem{Aspuru-Guzik12}
\bibinfo{author}{A. Aspuru-Guzik and P. Walther}, 
\newblock \bibinfo{title}{Photonic quantum simulators}, 
\newblock {\em Nat. Phys.}   \textbf{\bibinfo{volume}{8}}, 285-–291  (\bibinfo{year}{2012}).



\bibitem{Hong87}
\bibinfo{author}{C. K. Hong, Z. Y. Ou and L. Mandel,}
\newblock \bibinfo{title}{Measurement of subpicosecond time intervals between two photons by interference}, 
\newblock \bibinfo{journal}{Phys. Rev. Lett.}
\textbf{\bibinfo{volume}{59}}, 2044--2046
(\bibinfo{year}{1987}).




\bibitem{AA10}
\bibinfo{author}{S. Aaronson and A. Arkhipov,} 
\newblock In \emph{\bibinfo{booktitle}{Proceedings of the 43rd annual ACM symposium on Theory of computing}}
(\bibinfo{year}{ACM Press, 2011}).



\bibitem{Broome13boson}
\bibinfo{author}{M. A. Broome, A. Fedrizzi, S.  Rahimi-Keshari, J.  Dove, S.  Aaronson, T. C.  Ralph,  and A. G. White,  }
\newblock \bibinfo{title}{Photonic boson sampling in a tunable circuit}, 
\newblock \emph{\bibinfo{journal}{Science}}   \textbf{\bibinfo{volume}{339}}, \bibinfo{pages}{794–-798} (\bibinfo{year}{2013}).

\bibitem{Spring13boson}
\bibinfo{author}{J. B. Spring, B. J.  Metcalf, P.C.  Humphreys , W. S. Kolthammer, X. M.  Jin, M.  Barbieri, A.  Datta,   N. Thomas-Peter, N. K.   Langford, D.  Kundys, J. C. Gates,  B. J. Smith,  and I. A. Walmsley }, 
\newblock \bibinfo{title}{Boson Sampling on a photonic chip}, 
\newblock \emph{\bibinfo{journal}{Science}}   \textbf{\bibinfo{volume}{339}}, \bibinfo{pages}{798–-801} (\bibinfo{year}{2013}). 

\bibitem{Tillmann13boson}
\bibinfo{author}{ M. Tillmann, B.  Dakic, R.  Heilmann,   S. Nolte, A.  Szameit, and P. Walther, }
\newblock \bibinfo{title}{Experimental Boson Sampling},  
\newblock \emph{\bibinfo{journal}{Nat. Photon.}}   \textbf{\bibinfo{volume}{7}}, \bibinfo{pages}{540–-544} (\bibinfo{year}{2013}). 

\bibitem{Crespi13boson}
\bibinfo{author}{A. Crespi, R.  Osellame, R.  Ramponi, D. J. Brod, E. F.  Galv\~{a}o, N.  Spagnolo, C.  Vitelli, E.  Maiorino, P. Mataloni,  and  F. Sciarrino, }
\newblock \bibinfo{title}{Integrated multimode interferometers with arbitrary designs for photonic boson sampling},  
\newblock \emph{\bibinfo{journal}{Nat. Photon.}}   \textbf{\bibinfo{volume}{7}}, \bibinfo{pages}{545–-549} (\bibinfo{year}{2013}). 

\bibitem{Spagnolo13birthday}
\bibinfo{author}{N. Spagnolo, C. Vitelli, L. Sansoni, E. Maiorino, P. Mataloni, F. Sciarrino, D. J. Brod, E. F. Galv\~{a}o, A. Crespi, R. Ramponi and R. Osellame}, 
\newblock \bibinfo{title}{General Rules for Bosonic Bunching in Multimode Interferometers}, 
\newblock \emph{\bibinfo{journal}{Phys. Rev. Lett.}}   \textbf{\bibinfo{volume}{111}}, \bibinfo{pages}{130503} (\bibinfo{year}{2013}). 

\bibitem{Bentivegna15}
\bibinfo{author}{M. Bentivegna, N. Spagnolo, C. Vitelli, F. Flamini, N. Viggianiello, L. Latmiral, P. Mataloni, D.J. Brod, E. F. Galv\~{a}o, A. Crespi, R. Ramponi, R. Osellame and F. Sciarrino}, 
\newblock \bibinfo{title}{Experimental scattershot boson sampling}, 
\newblock \emph{\bibinfo{journal}{Sci. Adv.}}  \textbf{\bibinfo{volume}{1}}, \bibinfo{pages}{e1400255} (\bibinfo{year}{2015}). 

\bibitem{Carolan15}
\bibinfo{author}{J. Carolan, C. Harrold, C. Sparrow, E. Mart\'{i}n-L\'{o}pez, N. J.  Russell, J. W. Silverstone, P. J. Shadbolt, N. Matsuda, M. Oguma, M. Itoh, G.D. Marshall, M. G. Thompson , J. C. F. Matthews, T. Hashimoto, J. L. O'Brien and A. Laing}, 
\newblock \bibinfo{title}{Universal linear optics}, 
\newblock \emph{\bibinfo{journal}{Science}}   \textbf{\bibinfo{volume}{349}}, \bibinfo{pages}{711--716}  (\bibinfo{year}{2015}). 

\bibitem{Tillmann15}
\bibinfo{author}{M. Tillmann, S-H. Tan, S. E. Stoeckl, B. C. Sanders, H. de Guise, R. Heilmann, A. Szameit and P. Walther}, 
\newblock \bibinfo{title}{Generalized Multiphoton Quantum Interference}, 
\newblock \emph{\bibinfo{journal}{Phys. Rev. X}} \textbf{\bibinfo{volume}{5}}, \bibinfo{pages}{041015}  (\bibinfo{year}{2015}).




\bibitem{Wang_Pan17}
\bibinfo{author}{H. Wang, Y.  He, Y-H. Li, Z-E.  Su, B.	Li,	 H-L. Huang, X. Ding,	M-C. Chen, C.  Liu, J. Qin,	J-P. Li,	Y-M. He,	C. Schneider,	M. Kamp, C-Z. Peng,	S. Hofling, C-Y	 Lu, and J. W. Pan, }
\newblock \bibinfo{title}{High-efficiency multiphoton boson sampling}, 
\newblock \emph{\bibinfo{journal}{Nat. Photon.}}   \textbf{\bibinfo{volume}{11}}, \bibinfo{pages}{361--365}  (\bibinfo{year}{2017}). 

\bibitem{Loredo17}
\bibinfo{author}{J.C. Loredo, M.A. Broome, P. Hilaire, O. Gazzano, I. Sagnes, A. Lamaitre, M. P. Almeida, P. Senellart,  and A. G. White,  }
\newblock \bibinfo{title}{Boson Sampling with Single-Photon Fock States from a Bright Solid-State Source}, 
\newblock \emph{\bibinfo{journal}{Phys. Rev. Lett.}}   \textbf{\bibinfo{volume}{118}}, \bibinfo{pages}{130503}  (\bibinfo{year}{2017}). 

\bibitem{He_Pan17}
Y. He, X. Ding, Z-E. Su, H-L. Huang, J. Qin, C. Wang, S. Unsleber, C. Chen, H. Wang, Y-M. He, X-L. Wang, W-J. Zhang, S-J. Chen, C. Schneider, M. Kamp, L-X. You, Z. Wang, S. Höfling, C-Y. Lu and J. W. Pan,
\newblock Time-Bin-Encoded Boson Sampling with a Single-Photon Device,
\newblock {\em Phys. Rev. Lett.}  \textbf{118},  190501 (2017). 


\bibitem{Wang18}
\bibinfo{author}{H. Wang, W. Li, X. Jiang, Y. M. He, Y. H. Li, X. Ding, M. C. Chen, J. Qin, C. Z. Peng, C. Schneider, M. Kamp, W. J. Zhang, H. Li, L. X. You, Z. Wang, J. P. Dowling, S. H\"{o}fling, C. Y. Lu and J. W. Pan}
\newblock \bibinfo{title}{Toward scalable boson sampling with photon loss},
\newblock {\em preprint at arXiv:1801.08282} (\bibinfo{year}{2018}).

\bibitem{Wang18time}
\bibinfo{author}{X.-J. Wang, B. Jing, P.-F. Sun, C.-W. Yang, Y. Yu, V. Tamma, X.-H. Bao, J.-W. Pan}
\newblock \bibinfo{title}{Time-resolved boson sampling with photons of different colors},
\newblock {\em preprint at 	arXiv:1803.04696} (\bibinfo{year}{2018}).




\bibitem{Dakic12}
B. Dakic, Y. O. Lipp, X. Ma, M. Ringbauer, S. Kropatschek, S. Barz, T. Paterek, V. Vedral, A. Zeilinger, C. Brukner and P. Walther, 
\newblock Quantum discord as resource for remote state preparation,
\newblock {\em Nat. Phys.}  \textbf{8}, 666–-670 (2012). 


\bibitem{Vedral97}
V. Vedral, M. B. Plenio, M. A. Rippin, and P. L. Knight,
\newblock Quantifying Entanglement,
\newblock {\em Phys. Rev. Lett.}, \textbf{78}, 2275 (1997). 


\bibitem{Horodecki09}
R. Horodecki, P. Horodecki, M. Horodecki and K. Horodecki,
\newblock Quantum entanglement,
\newblock {\em Rev. Mod. Phys.}, \textbf{81} (2),  865--942 (2009). 

\bibitem{Aaronson14}
S. Aaronson and A. Arkhipov,
\newblock Bosonsampling is far from uniform,
\newblock {\em Quantum Inf. Comput.}, \textbf{14},  1383--1423 (2014). 

\bibitem{Tichy14}
\bibinfo{author}{M. C. Tichy, K. Mayer, A. Buchleitner and K. Molmer}, 
\newblock \bibinfo{title}{Stringent and Efficient Assessment of Boson-Sampling Devices}, 
\newblock \emph{\bibinfo{journal}{Phys. Rev. Lett.}}  \textbf{\bibinfo{volume}{118}}, \bibinfo{pages}{020502} (\bibinfo{year}{2014}). 

\bibitem{Aolita15}
\bibinfo{author}{L. Aolita, C. Gogolin, M. Kliesch and J. Eisert}, 
\newblock \bibinfo{title}{Reliable quantum certification of photonic state preparations}, 
\newblock \emph{\bibinfo{journal}{Nat. Commun.}}   \textbf{\bibinfo{volume}{6}}, \bibinfo{pages}{8948}  (\bibinfo{year}{2015}). 

\bibitem{Crespi15}
\bibinfo{author}{A. Crespi}, 
\newblock \bibinfo{title}{Suppression laws for multiparticle interference in Sylvester interferometers}, 
\newblock \emph{\bibinfo{journal}{Phys. Rev. A}}   \textbf{\bibinfo{volume}{91}}, \bibinfo{pages}{013811}  (\bibinfo{year}{2015}).

\bibitem{Walschaers16}
\bibinfo{author}{M. Walschaers, J. Kuipers, J. D. Urbina, K. Mayer, M. C. Tichy, K. Richter and A. Buchleitner}, 
\newblock \bibinfo{title}{Statistical benchmark for BosonSampling}, 
\newblock \emph{\bibinfo{journal}{New J. Phys.}}   \textbf{\bibinfo{volume}{18}}, \bibinfo{pages}{032001}  (\bibinfo{year}{2016}).

\bibitem{Bentivegna2016}
\bibinfo{author}{M. Bentivegna, N. Spagnolo and F. Sciarrino}, 
\newblock \bibinfo{title}{Is my boson sampling working?}, 
\newblock \emph{\bibinfo{journal}{New J. Phys.}}   \textbf{\bibinfo{volume}{18}}, \bibinfo{pages}{041001}  (\bibinfo{year}{2016}).

\bibitem{Wang2016}
\bibinfo{author}{S. T. Wang and L. M. Duan}, 
\newblock \bibinfo{title}{Certification of Boson Sampling Devices with Coarse-Grained Measurements}, 
\newblock \emph{\bibinfo{journal}{preprint at arXiv:1601.02627}}, (\bibinfo{year}{2016}).

\bibitem{Dittel2017}
\bibinfo{author}{C. Dittel, R. Keil and G. Weihs}, 
\newblock \bibinfo{title}{Many-body quantum interference on hypercubes}, 
\newblock \emph{\bibinfo{journal}{Quantum Sci. Technol.}}, \textbf{\bibinfo{volume}{2}}, \bibinfo{pages}{015003} (\bibinfo{year}{2017}).




\bibitem{Spagnolo14}
\bibinfo{author}{N. Spagnolo, C. Vitelli, M. Bentivegna, D. J. Brod, A. Crespi, F. Flamini, S. Giacomini, G. Milani, R. Ramponi, P. Mataloni, R. Osellame, E. F. Galv\~{a}o and F. Sciarrino}, 
\newblock \bibinfo{title}{Experimental validation of photonic boson sampling}, 
\newblock \emph{\bibinfo{journal}{Nat. Photon.}} \textbf{\bibinfo{volume}{8}}, \bibinfo{pages}{615--620} (\bibinfo{year}{2014}).


\bibitem{Carolan14}
\bibinfo{author}{J. Carolan, J. D. A. Meinecke, P. J. Shadbolt, N. J. Russell, N. Ismail, K. W\"{o}rhoff, T. Rudolph, M. G. Thompson, J. L. O'Brien, J. C. F. Matthews and A. Laing}, 
\newblock \bibinfo{title}{On the experimental verification of quantum complexity in linear optics}, 
\newblock \emph{\bibinfo{journal}{Nat. Photon.}} \textbf{\bibinfo{volume}{8}}, \bibinfo{pages}{621--626}  (\bibinfo{year}{2014}).


\bibitem{Bentivegna14}
\bibinfo{author}{M. Bentivegna, N. Spagnolo, C. Vitelli, D.J. Brod, A. Crespi, F. Flamini, R. Ramponi, P. Mataloni, R. Osellame, E. F. Galv\~{a}o and F. Sciarrino}, 
\newblock \bibinfo{title}{Bayesian approach to boson sampling validation}, 
\newblock \emph{\bibinfo{journal}{Int. J. Quantum Inf.}} \textbf{\bibinfo{volume}{12}}, \bibinfo{pages}{1560028} (\bibinfo{year}{2014}).


\bibitem{Crespi16}
\bibinfo{author}{A. Crespi, R. Osellame, R. Ramponi, M. Bentivegna, F. Flamini, N. Spagnolo, N. Viggianiello, L. Innocenti, P. Mataloni and F. Sciarrino}, 
\newblock \bibinfo{title}{Suppression law of quantum states in a 3D photonic fast Fourier transform chip}, 
\newblock \emph{\bibinfo{journal}{Nat. Commun.}} \textbf{\bibinfo{volume}{7}}, \bibinfo{pages}{10469}  (\bibinfo{year}{2016}).

\bibitem{Viggianiello17optimal}
\bibinfo{author}{N. Viggianiello, F. Flamini, M. Bentivegna, N. Spagnolo, A. Crespi, D. J. Brod, E. F. Galv\~{o}o, R. Osellame and F. Sciarrino}, 
\newblock \bibinfo{title}{Optimal photonic indistinguishability tests in multimode networks}, 
\newblock \emph{\bibinfo{journal}{preprint at arXiv:1710.03578}} (\bibinfo{year}{2017}).

\bibitem{Agresti17}
\bibinfo{author}{I. Agresti, N. Viggianiello, F. Flamini,  N. Spagnolo, A.
Crespi, R. Osellame, N. Wiebe and F. Sciarrino}, 
\newblock \bibinfo{title}{Pattern recognition techniques for Boson Sampling validation}, 
\newblock \emph{\bibinfo{journal}{preprint at arXiv:1712.06863}} (\bibinfo{year}{2017}).

\bibitem{Giordani18}
\bibinfo{author}{T. Giordani, F. Flamini, M. Pompili, N. Viggianiello, N. Spagnolo, A. Crespi, R. Osellame, N. Wiebe, M. Walschaers, A. Buchleitner and F. Sciarrino}, 
\newblock \bibinfo{title}{Experimental statistical signature of many-body quantum interference}, 
\newblock \emph{\bibinfo{journal}{Nat. Photon.}}  \textbf{\bibinfo{volume}{12}}, \bibinfo{pages}{173–-178}  (\bibinfo{year}{2018}).


\bibitem{Viggianiello18}
\bibinfo{author}{N. Viggianiello, F. Flamini, L. Innocenti, D. Cozzolino, M. Bentivegna, N. Spagnolo, A. Crespi, D. J. Brod, E. F. Galv\~{a}o, R. Osellame and F. Sciarrino},  
\newblock \bibinfo{title}{Experimental generalized quantum suppression law in Sylvester interferometers}, 
\newblock \emph{\bibinfo{journal}{New. J. Phys.}} \textbf{\bibinfo{volume}{20}},  \bibinfo{pages}{033017} (\bibinfo{year}{2018}).


\bibitem{Agne17}
\bibinfo{author}{S. Agne, T. Kauten, J. Jin, E. Meyer-Scott, J. Z. Salvail, D. R. Hamel, K. J. Resch, G. Weihs, and T. Jennewein}, 
\newblock \bibinfo{title}{Observation of Genuine Three-Photon Interference}, 
\newblock \emph{\bibinfo{journal}{Phys. Rev. Lett.}}  \textbf{\bibinfo{volume}{118}}, \bibinfo{pages}{153602}  (\bibinfo{year}{2017}).


\bibitem{Menssen15}
\bibinfo{author}{A. J. Menssen, A. E. Jones, B. J. Metcalf, M. C. Tichy, S. Barz, W. S. Kolthammer, and I. A. Walmsley}, 
\newblock \bibinfo{title}{Distinguishability and Many-Particle Interference}, 
\newblock \emph{\bibinfo{journal}{Phys. Rev. Lett.}} \textbf{\bibinfo{volume}{118}}, \bibinfo{pages}{153603}  (\bibinfo{year}{2017}).

\bibitem{Osellame2003}
\bibinfo{author}{R. Osellame, S. Taccheo, M. Marangoni, R. Ramponi, P. Laporta, D. Polli, S. De Silvestri and G. Cerullo}, 
\newblock \bibinfo{title}{Femtosecond writing of active optical waveguides with astigmatically shaped beams},
\newblock \emph{\bibinfo{journal}{J. Opt. Soc. Am. B}} \textbf{\bibinfo{volume}{20}}, \bibinfo{pages}{1559–-1567}  (\bibinfo{year}{2003}).

\bibitem{obrien09}
\bibinfo{author}{G. D. Marshall, A. Politi, J. C. F. Matthews, P. Dekker, M. Ams, M. J. Withford and J. L. O’Brien}, 
\newblock \bibinfo{title}{Laser written waveguide photonic quantum circuits}, 
\newblock \emph{\bibinfo{journal}{Opt. Exp.}} \textbf{\bibinfo{volume}{17}}, \bibinfo{pages}{12546--12554}  (\bibinfo{year}{2009}).

\bibitem{szamait15}
\bibinfo{author}{T. Meany, M. Gr\"{a}fe, R. Heilmann, A. Perez-Leija, S. Gross, M. J. Steel, M. J. Withford and A. Szameit,}
\newblock \bibinfo{title}{Laser written circuits for quantum photonics}, 
\newblock \emph{\bibinfo{journal}{Laser and photonics reviews}} \textbf{\bibinfo{volume}{9}},  \bibinfo{pages}{363--384} (\bibinfo{year}{2015}).

\bibitem{Tanzilli}
\bibinfo{author}{S. Tanzilli, A. Martin, F. Kaiser, M. De Micheli, O. Alibart, and D. B. Ostrowsky}, 
\newblock \bibinfo{title}{On the Genesis and Evolution of Integrated Quantum Optics}, 
\newblock {\em Laser and Photonics Reviews}   \textbf{\bibinfo{volume}{6}}, 115–-143  (\bibinfo{year}{2012}).

\bibitem{barak}
\bibinfo{author}{R. Barak and Y. Ben-Aryeh}, 
\newblock \bibinfo{title}{Quantum fast fourier transform and quantum computation by linear optics}, 
\newblock \emph{\bibinfo{journal}{J. Opt. Soc. Am. B}} \textbf{\bibinfo{volume}{24}},  \bibinfo{pages}{231–-240}  (\bibinfo{year}{2007}). 




\bibitem{Rohde}
\bibinfo{author}{P. Rohde}, 
\newblock \bibinfo{title}{Boson sampling with photons of arbitrary spectral structure}, 
\newblock \emph{\bibinfo{journal}{Phys. Rev. A}} \textbf{\bibinfo{volume}{91}}, \bibinfo{pages}{012307}  (\bibinfo{year}{2015}).

\bibitem{Tichy}
\bibinfo{author}{M. Tichy}, 
\newblock \bibinfo{title}{Sampling of partially distinguishable bosons and the relation to the multidimensional permanent},
\newblock \emph{\bibinfo{journal}{Phys. Rev. A}} \textbf{\bibinfo{volume}{91}}, \bibinfo{pages}{022316}  (\bibinfo{year}{2015}).


\bibitem{Scheschnovich}
\bibinfo{author}{V. Shchesnovich}, 
\newblock \bibinfo{title}{Partial indistinguishability theory for multiphoton experiments in multiport devices}, 
\newblock \emph{\bibinfo{journal}{Phys. Rev. A}} \textbf{\bibinfo{volume}{91}}, \bibinfo{pages}{013844}  (\bibinfo{year}{2015}).


\bibitem{footnote1} The identification of $|A \cap B|$ with \unexpanded{$\left| \braket{A}{B} \right| ^2$} assumes each photon is in a pure state. However, the decomposition of Eq.\ (\ref{eq:convexn}) allows for arbitrary mixed states inconsistent with a pure state for each photon. In this case, one can identify $|A \cap B|$ with $\sum_i c_i$, where the sum runs over all extremal states in Eq.\ (\ref{eq:convexn}) where photons $A$ and $B$ are identical. Equivalently, we can identify it with $(2 p_{AB} -1)$, where $p_{AB}$ is the bunching probability of a HOM test between photons $A$ and $B$.

\bibitem{supp} In Supplemental Material we report the proof of the theorem for the interpretation in set theory, experimental details of our setup, a discussion on the analysis performed considering multi-pair emission provided by PDC sources and a study of the robustness of the protocol in presence of non-ideal reflectivities.

\bibitem{Garcia-Escartin}
\bibinfo{author}{J. C. Garcia-Escartin and P. Chamorro-Posada}, 
\newblock \bibinfo{title}{SWAP test and Hong-Ou-Mandel effect are equivalent}, 
\newblock \emph{\bibinfo{journal}{Phys. Rev. A}} \textbf{\bibinfo{volume}{87}}, \bibinfo{pages}{052330}  (\bibinfo{year}{2013}).

\end{thebibliography}
\end{document}